\documentclass[aps,prl,twocolumn,showpacs,nobibnotes,10pt]{revtex4-1}
\usepackage{graphics,graphicx,amsfonts,amsmath,amsbsy,amssymb,color}
\usepackage{mathrsfs}
\usepackage{multirow}
\usepackage{subdepth}
\usepackage{bm}
\usepackage{braket}
\usepackage{subfigure}
\usepackage{xspace}

\newcommand{\ve}[1]{\ensuremath{{\mathbf #1}}\xspace}

\begin{document}
\author{Robert~E.~Thomas$^{(a)}$}
\author{Daniel~Opalka$^{(a,b)}$}
\author{Catherine~Overy$^{(a)}$}
\author{Peter~J.~Knowles$^{(c)}$}
\author{Ali~Alavi$^{(a,b)}$}
\email{asa10@cam.ac.uk}
\author{George~H.~Booth$^{(a,d)}$}
\email{george.booth@kcl.ac.uk}
\affiliation{$^{(a)}$University of Cambridge, The University Chemical Laboratory, Lensfield Road, Cambridge CB2 1EW, U.K.}
\affiliation{$^{(b)}$Max Planck Institute for Solid State Research, Heisenbergstra{\ss}e 1, 70569 Stuttgart, Germany}
\affiliation{$^{(c)}$Cardiff University, School of Chemistry, Cardiff CF10 3AT, U.K.}
\affiliation{$^{(d)}$King's College London, Department of Physics, Strand, London WC2R 2LS, U.K.}
\title{Analytic nuclear forces and molecular properties from full configuration interaction quantum Monte Carlo}


\begin{abstract}
Unbiased stochastic sampling of the one- and two-body reduced density matrices is achieved in full configuration interaction quantum Monte Carlo with the introduction of a second, ``replica'' ensemble of walkers, whose population evolves in imaginary time independently from the first, and which entails only modest additional computational overheads. The matrices obtained from this approach are shown to be representative of full configuration-interaction quality, and hence provide a realistic opportunity to achieve high-quality results for a range of properties whose operators do not necessarily commute with the hamiltonian. A density-matrix formulated quasi-variational energy estimator having been already proposed and investigated, the present work extends the scope of the theory to take in studies of analytic nuclear forces, molecular dipole moments and polarisabilities, with extensive comparison to exact results where possible. These new results confirm the suitability of the sampling technique and, where sufficiently large basis sets are available, achieve close agreement with experimental values, expanding the scope of the method to new areas of investigation.
\end{abstract}
   

\maketitle

\section{Introduction}

The full configuration interaction quantum Monte Carlo method (FCIQMC) and its initiator adaptation ($i$-FCIQMC) are projector QMC techniques, capable of providing near-exact, systematically improvable descriptions of correlated wavefunctions expressed as linear combinations of Slater determinants.\cite{booththomalavi2009,clelandboothalavi2010} This convergence is achieved by stochastically sampling the exponentially large (though finite) Hilbert spaces of configuration interaction theory \emph{via} a population dynamics performed on an ensemble of signed walkers. Annihilation processes provide a means of combating the ill effects of the fermion sign problem which plagues projector approaches,\cite{spencerbluntfoulkes2012,shepherdscuseriaspencer2014,pederivakalos2011} exploiting the sparsity of the wavefunction induced by a judicious choice of orbital basis.
The approach requires substantially less computational effort than an iterative diagonalisation technique, and has thus found considerable success in studies of atomic and molecular systems,\cite{boothalavi2010,clelandboothalavi2011,boothclelandthomalavi2011,clelandboothoveryalavi2012,daday2012,thomasoveryboothalavi2014,thomasboothalavi2015} model systems such as the homogeneous electron gas and the Hubbard model,\cite{shepherdboothalavi2012,shepherdboothgrueneisalavi2012,schwarzboothalavi2015} and solid-state systems.\cite{boothgrueneiskressealavi2013}

The principal focus of many of these studies has been to derive properties based upon total energies, for which an unbiased projected estimator is readily available, and which have included excitation and dissociation energies,\cite{boothclelandthomalavi2011,clelandboothoveryalavi2012,daday2012} electron affinities,\cite{clelandboothalavi2011} ionisation potentials,\cite{boothalavi2010,thomasboothalavi2015} and equations of state.\cite{boothgrueneiskressealavi2013}
Despite their success, however, the extension to include the calculation of a greater range of properties --- expectation values of operators which do not necessarily commute with the hamiltonian --- remains highly desirable. This focus has been the subject of considerable interest for the QMC community in general, and has posed a considerable challenge for decades.\cite{east1988,reynolds1990,barnett1991,barnett1992,langfelder1997,baroni1999,caffarel2000,caffarel2003,gaudoin2007,gaudoin2010,sorella2010,rothstein2013,overy2014} These properties, which include static correlation functions and entropy estimators as well as the forces, multipole moments, and polarisabilities considered here, may be deduced from the effect of a perturbation from the corresponding operator, $\hat{P}$, upon the hamiltonian,
\begin{equation}
\hat{H}^\prime = \hat{H} + \lambda\hat{P},
\end{equation}
with $\lambda$ the perturbation strength, such that the expectation value, $\langle\hat{P}\rangle$, is given by the derivative of the energy with respect to $\lambda$, evaluated at $\lambda=0$:
\begin{equation}
\langle\hat{P}\rangle=\left.\frac{\partial E\left(H^\prime\right)}{\partial\lambda}\right\vert_{\lambda=0}.
\end{equation} 
In accordance with the Hellmann--Feynman theorem,\cite{feynman1939} applicable to converged (normalised) \mbox{$i$-FCIQMC} wavefunctions by analogy with deterministic and strictly variational FCI, this expression reduces to
\begin{equation}
\langle\hat{P}\rangle=\langle \Psi | \hat{P} | \Psi \rangle,
\end{equation}
or equivalently to the trace of $\hat{P}$ with the appropriate rank of reduced density matrix.\cite{mazziotti2007} It is worth noting that unconverged $i$-FCIQMC wavefunctions need not rigorously obey the Hellmann--Feynman theorem, and so in this work we ensure that we are working in the large walker limit, such that systematic errors in the sampled distribution due to insufficient walker numbers have been minimized to the FCI-limit. 

The effective stochastic acquisition of these reduced density matrices, therefore, has the capacity to broaden the scope of \mbox{$i$-FCIQMC} significantly, and motivates the present work.
We begin with a brief overview of the $i$-FCIQMC algorithm, including its extension to non-integer walker weights,\cite{umrigar2012} before recapitulating some of the details of the ``replica'' density-matrix sampling technique.\cite{zhangkalos1993,overy2014,bluntalavibooth2015} Building upon that previous work, our discussion turns to consider the calculation of nuclear forces, molecular dipole moments, and atomic dipole polarisabilities, and in so doing confirms the high quality of the sampled one- and two-body reduced density matrices which is now achievable. 

\section{Methodology}

\subsection{$i$-FCIQMC}

Initiator full configuration interaction quantum Monte Carlo provides stochastic integration of the $N$-electron, imaginary-time Schr\"{o}dinger equation, yielding wavefunctions expressed as a linear combination of the set of Slater determinants, $\left\{\Ket{D_\mathbf{i}}\right\}$, formed from the underlying one-particle (most often Hartree--Fock) basis:
\begin{equation}
\Psi = \sum_\mathbf{i} C_\mathbf{i}\Ket{D_\mathbf{i}}.
\label{fci-expansion}
\end{equation}
The coefficients of this wavefunction expansion are obtained by iterative application of the equations
\begin{equation}
C_\mathbf{i}\left(\tau + \delta\tau\right) = C_\mathbf{i}\left(\tau\right)-\delta\tau\left(H_\mathbf{ii} - \mu\right)C_\mathbf{i}\left(\tau\right) -\sum_{\mathbf{j}\neq\mathbf{i}}\delta\tau H_{\mathbf{ij}}C_\mathbf{j}\left(\tau\right),
\label{evolution}
\end{equation}
representing the evolution of the coefficients over a timestep $\delta\tau$ in imaginary time. This evolution is achieved by subjecting an ensemble of signed walkers to a three-step population dynamics algorithm of ``spawning'', ``death'', and ``annihilation'' steps, the walker populations, $\left\{N_\mathbf{i}\right\}$, becoming proportional to the coefficients. The full details of this approach have been expounded in previous papers,\cite{booththomalavi2009,clelandboothalavi2010,boothalavi2010,boothclelandthomalavi2011,clelandboothoveryalavi2012,boothgrueneiskressealavi2013,thomasoveryboothalavi2014,boothsmartalavi2014,thomasboothalavi2015}, and what follows should be regarded only as a brief summary. 

Typically initialised with a single walker placed upon the Hartree--Fock determinant, $\Ket{D_\mathbf{0}}$, a simulation using integer walkers proceeds with a coupled determinant, $\Ket{D_\mathbf{j}}$, being randomly selected for each walker on parent determinant, $\Ket{D_\mathbf{i}}$, with a probability $p_\mathrm{gen}\left(D_\mathbf{j}\vert D_\mathbf{i}\right)$. The determinant selected, the parent walker then attempts to spawn a child on to it, with a probability
\begin{equation}
p_\mathrm{s}\left(D_\mathbf{j}\vert D_\mathbf{i}\right) = \frac{\delta\tau\left\vert H_\mathbf{ij} \right\vert}{p_\mathrm{gen}\left(D_\mathbf{j}\vert D_\mathbf{i}\right)}.
\end{equation}
If the attempt is successful, the sign of the spawned walker matches that of its parent if $H_\mathbf{ij}<0$ and is inverted if $H_\mathbf{ij}>0$. The initiator adaptation, $i$-FCIQMC, modifies the spawning step by introducing a parameter, $n_\mathrm{a}$, which specifies a lower population threshold under which a parent determinant is prevented from spawning on to unoccupied determinants.
Each walker next attempts to die, with a probability given by
\begin{equation}
p_\mathrm{d}\left(D_\mathbf{i}\right) = \delta\tau\left(H_\mathbf{ii} - \mu\right),
\end{equation}
in which $\mu$ is a population control parameter --- known as the ``shift'' --- which tends to the ground-state energy in the long-$\tau$ limit. 

These two steps are themselves sufficient to describe Eq.~\ref{evolution} fully, but are insufficient to provide convergence to a fermionic wavefunction. Instead, a third step --- ``annihilation'' --- is required in order to suppress the deleterious effects of the fermion sign problem.\cite{spencerbluntfoulkes2012,kolodrubetz2013} After each iteration, walkers of opposite sign on the same determinant annihilate, and in so doing ensure that each determinant is populated by walkers of only one sign for the next iteration. The success of these processes relies on the sparsity of the wavefunction induced by the underlying basis --- typically chosen to be Hartree--Fock orbitals --- which confines it to a generally small region of the Hilbert space. In so doing, it ensures that annihilation events are numerous enough to maintain the sign structure of the sampled wavefunction accurately. 

Although the walkers of ($i$)-FCIQMC were initially conceived as an ensemble of discrete particles, there is some merit in instead positing a set of non-integer walkers.\cite{umrigar2012,overy2014} Such an approach reduces the amount of random number generation required, reduces the instantaneous fluctuations in the populations on a given determinant, and hence the fluctuations in the energy estimators in imaginary time.

This formulation is achieved by applying the spawning, death, and annihilation steps introduced earlier \emph{continuously}, rather than discretely. Thus, instead of spawning a walker of signed integer weight from a determinant $\Ket{D_\mathbf{i}}$ to a coupled determinant $\Ket{D_\mathbf{j}}$ with a probability $p_\mathrm{s}\left(D_\mathbf{j}\vert D_\mathbf{i}\right)$, a walker of weight $p_\mathrm{s}$ is spawned with probability $1$. Likewise, the death step is remodelled such that it simply involves reducing the population on a determinant $\Ket{D_\mathbf{i}}$ by $p_\mathrm{d}\left(D_\mathbf{i}\right)$. Annihilation is achieved by taking the signed sum of walkers on each determinant on a given iteration as the residual population for the next iteration. For $i$-FCIQMC calculations, the parameter $n_\mathrm{a}$ is recast as a continuous variable rather than an integer.

The continuous nature of the spawned walkers in this approach does not, however, imply that the number of spawning \emph{events} becomes continuous. As in the integer formulation, where there are exactly $N_\mathbf{i}$ spawning attempts from determinant $\Ket{D_\mathbf{i}}$ with a population $N_\mathbf{i}$ on each iteration, there are a discrete number of attempts per determinant per iteration. For practical purposes, a continuous spawning threshold, $\kappa$, is introduced such that if $p_\mathrm{s} < \kappa$, $\kappa$ walkers are spawned with a probability $p_\mathrm{s}/\kappa$. This implementation is designed to alleviate the significant cost of low-weighted spawnings compared to their effect on the overall wavefunction, as well as ensuring that the wavefunction remains compact and expressible instantaneously by a number of walkers far smaller than the size of the space. 

Whilst the death step requires no extra modification of this kind, some additional considerations must be addressed for annihilation. In order that determinants can become completely depopulated, and we are not forced to store large numbers whose populations are very close to, but not exactly, zero, a minimum occupation threshold, $N_\mathrm{occ}$, is imposed upon them. If, after annihilation, the population on a determinant $N_\mathbf{i}<N_\mathrm{occ}$, its population is set either to $N_\mathrm{occ}$ with probability $N_\mathbf{i}/N_\mathrm{occ}$, or else to $0$ with probability $1-N_\mathbf{i}/N_\mathrm{occ}$.

As a final practical means of alleviating the computational burden of this approach, it is possible to treat only a subspace of the full Hilbert space with non-integer walkers, continuing to describe the remainder in a discretised fashion. In order to preserve the benefits of the non-integer approach on the fluctuations of the energy estimators, the truncation is specified by an excitation level, $\chi$, with only $\chi$-fold and lower excitations from the reference included in the non-integer subspace. A typical choice of parameters $N_\mathrm{occ}=1$, $2\leq\chi\leq4$ ($4$ is used here), and $\kappa=0.01$ entails only a modest increase in the computational cost of the calculation over the integer implementation, while retaining many of the benefits of the full non-integer approach.

$i$-FCIQMC provides two essentially independent energy estimators, which, taken together, provide a useful confirmation of the validity of the obtained result. The first, to which we have already alluded, is the shift, $\mu$. This is initially held constant (typically at zero) to facilitate an exponential growth in the number of walkers, before being allowed to vary dynamically to keep the population constant. At convergence, this variation fluctuates around the energy of system, and thus provides an energy on the basis of the growth rate of the entire ensemble of walkers. A projected energy estimator, of the form
\begin{equation}
E_{\mathrm{proj}}\left(\tau\right) = \frac{\langle D_\mathbf{0} | \hat{H} | \Psi \left(\tau\right) \rangle }{\Braket{D_\mathbf{0}|\Psi\left(\tau\right)}},
\end{equation}
on the other hand, depends only upon the populations of the determinants coupled to the reference state, $\Ket{D_\mathbf{0}}$. Whilst the error in this projection is formally first-order in the wavefunction error, its non-variationality tends to mean that it converges rather faster to the exact, infinite-walker limit than does a variational estimator, owing to favourable cancellation of errors.
The projected energy is thus typically preferred when the wavefunction is dominated by the Hartree--Fock determinant, but a projection on to a multi-reference trial wavefunction or the variational estimator provided by the density matrices (which is second-order in the wavefunction error) are often more useful in more strongly-correlated cases.\cite{overy2014} Once the ensemble has equilibrated, the simulation is allowed to evolve in imaginary time until the statistical errors in both $\mu$ and $E_\mathrm{proj}$ have been satisfactorily reduced, upon which a Flyvbjerg--Petersen blocking analysis is performed to estimate the error in the obtained result.\cite{flyvbjerg1989}

\subsection{Stochastic density-matrix sampling}

In terms of the wavefunction ansatz of $i$-FCIQMC (Eq.~\ref{fci-expansion}) and the creation and annihilation operators, the one- and two-body reduced density matrices, $\bm{\gamma}$ and $\bm{\Gamma}$, may be formulated in terms of the wavefunction expansion and the conventional creation and annihilation operators, $\hat{a}^\dagger$ and $\hat{a}$, as
\begin{align}
\gamma_{pq} &= \langle \Psi|\hat{a}_p^\dagger\hat{a}_q|\Psi \rangle\\
            &=\sum_{\mathbf{ij}}C_\mathbf{i}C_\mathbf{j}\langle D_\mathbf{i}|\hat{a}_p^\dagger\hat{a}_q|D_\mathbf{j} \rangle,
\end{align}
and,
\begin{align}
\Gamma_{pqrs} &= \langle \Psi|\hat{a}_p^\dagger\hat{a}_q^\dagger\hat{a}_s\hat{a}_r|\Psi \rangle\\
              &= \sum_{\mathbf{ij}}C_{\mathbf{i}}C_{\mathbf{j}}\langle D_{\mathbf{i}}|\hat{a}_p^\dagger\hat{a}_q^\dagger\hat{a}_s\hat{a}_r|D_{\mathbf{j}} \rangle,
\end{align}
respectively,\cite{blum,helgaker} and an important recent development of the theory allows these objects to be sampled in an efficient, stochastically unbiased fashion.\cite{boothclelandalavitew2012,overy2014}

The diagonal elements of these objects, of the form
\begin{equation}
\Gamma_{pqpq} = \sum_{\mathbf{i}\ni\left\{p,q\right\}} C_\mathbf{i}^2,
\end{equation}
may be calculated straightforwardly, as each determinant, $\Ket{D_\mathbf{i}}$, contributes $C_\mathbf{i}^2$ to each of the $\frac{N\left(N-1\right)}{2}$ matrix elements involving its occupied orbitals.
The corresponding explicit generation of all the required determinant pairs for the off-diagonal elements is not practical, but the observation that the relevant pairs are at most double excitations of one another allows both $\bm{\gamma}$ and $\bm{\Gamma}$ to be sampled \emph{via} the spawning steps.\cite{boothclelandalavitew2012} Thus, the existing computational effort required for the communication of the spawning event need only be slightly accentuated (by the need now to convey both the amplitude and the identity of the parent determinant to the child) to allow the contributions to the off-diagonal matrix elements from determinant pairs to be calculated on the fly.

As these off-diagonal contributions are only added upon a \emph{successful} spawning event, it is necessary that they be rescaled according to the probability of such an event taking place. That is, a contribution $C_\mathbf{i}C_\mathbf{j}$ will instead be accumulated as
\begin{equation}
    \frac{C_\mathbf{i}C_\mathbf{j}\Braket{D_\mathbf{i}|\hat{a}_p^\dagger\hat{a}_q^\dagger\hat{a}_s\hat{a}_r|D_\mathbf{j}}}{p_\mathrm{c}\left(D_\mathbf{j}\vert D_\mathbf{i}\right)},
\end{equation}
with $p_\mathrm{c}\left(D_\mathbf{j}\vert D_\mathbf{i}\right)$ the probability that at least one spawning attempt from $\Ket{D_\mathbf{i}}$ to $\Ket{D_\mathbf{j}}$ is successful on a given iteration. Depending on whether integer or non-integer walkers are considered, this probability is given by
\begin{align}
p_\mathrm{c} =
\begin{cases}
1-\lambda^{N_\mathbf{i}} & N_\mathbf{i}\in\mathbb{Z}\\
1 - \left(\left\lceil N_\mathbf{i}\right\rceil - N_\mathbf{i}\right)\lambda^{\left\lfloor N_{\mathbf{i}}\right\rfloor} - \left(N_\mathbf{i} - \left\lfloor N_\mathbf{i}\right\rfloor\right)\lambda^{\left\lceil N_\mathbf{i} \right\rceil} & N_\mathbf{i}\notin\mathbb{Z},\\
\end{cases}
\end{align}
with $N_\mathbf{i}$ the instantaneous walker population residing on $\Ket{D_\mathbf{i}}$ and $\lambda$ the probability that no walker is spawned between $\Ket{D_\mathbf{i}}$ and $\Ket{D_\mathbf{j}}$ in a single attempt. For an integer spawning event, this probability is
\begin{equation}
\lambda_{\mathrm{int}} = 1 - \min\left(\delta\tau\left\vert H_{\mathbf{ij}}\right\vert, p_\mathrm{gen}\left(D_\mathbf{j}\vert D_\mathbf{i}\right)\right),
\end{equation}
but this must be modified in the case of continuous spawning to
\begin{align}
\lambda_{\mathrm{cont}} =
\begin{cases}
1 - \frac{\delta\tau\left\vert H_{\mathbf{ij}}\right\vert}{\kappa} & p_\mathrm{s} < \kappa\\
1 - p_\mathrm{gen}\left(D_\mathbf{j}\vert D_\mathbf{i}\right) & \mathrm{otherwise},\\
\end{cases}
\end{align}
where $\kappa$ is the continuous spawning threshold, if used.

A na\"ive implementation of the above sampling is satisfactory for the accumulation of approximate density matrices, but is beset by a number of shortcomings which should be considered.\cite{overy2014} As contributions to the off-diagonal matrix elements are only added upon a successful spawning attempt, problems can arise when the spawning events are discretised. In this case, the probability that such an event occurs is proportional to the coupling hamiltonian matrix element, $H_\mathbf{ij}$, and pairs of determinants which are connected by large matrix elements are correspondingly sampled more often than pairs which are only weakly coupled. Thus, if two highly-weighted determinants contained in the stochastically-sampled, integer walker space are connected by a small hamiltonian element, their contribution to the density matrices may be severely under-represented, or neglected entirely.

This problem is most notably in evidence in the case of single excitations of the reference determinant, for which the coupling matrix elements are strictly zero according the Brillouin's theorem. This is countered in the present implementation by accounting for these contributions to the density matrices explicitly, and hence removing the dependence upon a successful spawning event. Other contributions, however, whose sampling will still be proportional to the reduced hamiltonian,\cite{mazziotti2012} defined in terms of the one- and two-electron integrals, $\left\{h_{pq}\right\}$ and $\left\{g_{pqrs}\right\}$, as
\begin{equation}
k_{pqrs} = \frac{1}{2N-2}(h_{pr}\delta_{qs} + h_{qs}\delta_{pr}) + g_{pqrs},
\end{equation}
will give rise to an biasing error in density matrices in the long-$\tau$ limit for determinant pairs where $k_{pqrs}\approx0$, but whose amplitudes are both significantly non-zero. However, modifications to the algorithm to treat the bias remaining beyond that already defined by Brillouin's theorem explicitly --- such as introducing additional events to spawn walkers proportionally to the inverse of the hamiltonian element --- have been shown to be of little additional benefit due to the negligible nature of this bias in numerical studies to date.\cite{overy2014} 

In a separate difficulty, it has been shown previously that a straightforward implementation of the above sampling gave rise to a convergence of the density matrices with increasing $N_\mathrm{w}$ which was rather slower than that of, say, the projected energy. This behaviour stems not simply from undersampling, but rather from a bias in the statistical sampling technique itself. In particular, appropriate contributions to the matrix elements are approximated by
\begin{align}
\langle N_\mathbf{i}\left(\tau\right)\rangle_\tau \langle N_\mathbf{j}\left(\tau\right)\rangle_\tau &= \langle N_\mathbf{i}(\tau)N_\mathbf{j}(\tau)\rangle_\tau - \sigma(N_\mathbf{i}(\tau),N_\mathbf{j}(\tau))\\
&\approx \left\langle N_\mathbf{i}\left(\tau\right)N_\mathbf{j}\left(\tau\right)\right\rangle_\tau,
\end{align}
ignoring the potentially significant covariance, $\sigma$, between the two amplitudes and introducing a bias, whether or not the averaged walker populations are themselves unbiased. It is, to that end, unsurprising that this problem is at its greatest for diagonal elements, for which the ``two'' amplitudes are perfectly correlated, and --- the error being of a single sign --- there is no possibility of error cancellation.

This problem is rather more serious than the previous concerns over discretised spawning, but one for which a rather simple solution exists. Unbiased density matrices can be calculated with the introduction of a second, uncorrelated walker ensemble, to which the stochastic spawning, death, and annihilation steps are applied independently, and whose statistics are acquired separately, from the first.\cite{overy2014} This adaptation, known as replica sampling, achieves the unbiasing by ensuring that all the products of determinant amplitudes are calculated using populations from both simulations, and has previously found application in the stochastic sampling of the $N$-electron density matrix known as density matrix quantum Monte Carlo,\cite{blunt2014} and the recently-introduced Krylov-projected quantum Monte Carlo.\cite{bluntalavibooth2015} That is, for example, a successful spawning event from $\Ket{D_\mathbf{i}}$ to $\Ket{D_\mathbf{j}}$ in replica $1$, occurring with a probability $p_\mathrm{c}^{\left(1\right)}\left(D_\mathbf{j}\vert D_\mathbf{i}\right)$, gives rise to a contribution of:
\begin{equation}
\frac{N_\mathbf{i}^{\left(1\right)}N_\mathbf{j}^{\left(2\right)}}{p_\mathrm{c}^{\left(1\right)}\left(D_\mathbf{j}\vert D_\mathbf{i}\right)} + \frac{N_\mathbf{i}^{\left(2\right)}N_\mathbf{j}^{\left(1\right)}}{p_\mathrm{c}^{\left(2\right)}\left(D_\mathbf{j}\vert D_\mathbf{i}\right)}.
\end{equation}
This approach bears some conceptual similarity with the bilinear sampling algorithm in Green's function Monte Carlo, introduced by Zhang and Kalos, in that both seek a means of finding expectation values of operators which do not commute with the hamiltonian, \emph{via} two sets of independent walker distributions.\cite{zhangkalos1993} The main difference, though, is that the bilinear approach transforms the Schr\"{o}dinger equation such that there are two related wavefunctions to sample, while in the present work the walker ensembles are independent samples of the same underlying object.
In providing a stochastically unbiased route to the density matrices, the replica sampling technique thus provides the first realistic opportunity to achieve high-accuracy \emph{ab initio} results for the sizeable suite of properties that can be derived therefrom.

\section{Nuclear forces}
The force acting on a nucleus in a molecule or cluster is defined as the
negative gradient of the molecular energy with respect to the nuclear
coordinates:
\begin{equation}
    \ve{F} = -\frac{\partial E}{\partial \ve{R}}.
    \label{eq:force}
\end{equation}
In Eq.~\eqref{eq:force} the symbol $\ve{F}$ denotes the nuclear force
vector, $E$ the energy of a molecule at a fixed geometry in the electronic
ground state, and $\ve{R}$ refers to the vector of nuclear coordinates in the centre of mass frame of reference.
A comprehensive review of techniques and explicit expressions to compute
derivatives of the electronic energy with respect to nuclear coordinates is
available in the literature.\cite{kato1979,knowlessextonhandy1982,Yamaguchi1994} The following discussion is thus limited to the basic concepts for the
calculation of nuclear forces using all-electron FCI wavefunctions as 
obtained as a statistical average using the \mbox{$i$-FCIQMC} method, once the
calculation has been converged with respect to the number of walkers.
In the present work, we have adjusted the total number of walkers to
achieve a walker population of 50000 at the reference (i.\,e.\ highest
populated) determinant.
Preceding work confirmed that, at such a population levels, noise arising from
small stochastic populations of random determinants is sufficiently
suppressed and the wavefunction converged.

The first derivative of the electronic energy of a CI wavefunction
generally depends on the derivatives of the atomic orbitals (AOs) and the
molecular-orbital (MO) and CI coefficients.
All these terms depend upon the nuclear coordinates, and the computation of
nuclear forces requires knowledge of the first derivatives with
respect to all considered degrees of freedom.
However, electronic wave functions obtained from $i$-FCIQMC optimizations are
variational with respect to the CI coefficients and a component $\ve{F}_x$ of
the nuclear force vector can be expressed in terms of the reduced
density matrices as
\begin{equation}
    \ve{F}_x = -\sum^{\text{MO}}_{pq}\gamma_{pq}\frac{\partial h_{pq}}{\partial x} 
        - \frac{\partial h_{\text{nu}}}{\partial x}
        - \sum^{\text{MO}}_{pqrs}\Gamma_{pqrs}\frac{\partial\left( pq|rs \right)}{\partial x},
    \label{eq:fx}
\end{equation}
in which the terms $\left\{h_{pq}\right\}$ represent the one-electron integrals from the hamiltonian, and $h_{\mathrm{nu}}$ is the contribution from the fixed nuclei.
Moreover, all-electron FCI wavefunctions considered in this work are
also invariant under variation of the MO coefficients.
The nuclear forces can thus be expressed solely in terms of the one- and
two-electron density matrices and the skeleton derivative integrals of the
basis functions:
\begin{equation}
    \begin{split}
        \ve{F}_x& = -\sum^{\text{MO}}_{pq}\sum^{\text{AO}}_{\mu\nu}\gamma_{pq}
            C_{\mu p}C_{\nu q}\frac{\partial h_{\mu\nu}}{\partial x}
            - \frac{\partial h_{\text{nu}}}{\partial x} \\ 
                & -\sum^{\text{MO}}_{pqrs}\sum^{\text{AO}}_{\mu\nu\rho\sigma}\Gamma_{pqrs}C_{\mu
                    p}C_{\nu q}C_{\rho r}C_{\sigma s}\frac{\partial \left(
                    \mu\nu|\rho\sigma \right)}{\partial x}  \\
                & + \sum^{\text{MO}}_{pq}\sum^{\text{AO}}_{\mu\nu}
                    X_{pq}C_{\mu p}C_{\nu q}\frac{\partial S_{\mu\nu}}{\partial x},
    \end{split}
\end{equation}
where
\begin{equation}
    X_{pq} = \sum_{r}^{\text{MO}}\gamma_{pr}h_{qr} +
        2\sum_{rst}^{\text{MO}}\Gamma_{prst}\left( qr|st \right)
    \label{eq:x}
\end{equation}
is an element of the lagrangian and $S_{\mu\nu}$ is an element of the overlap
matrix.
In particular, neither the computation of derivatives of the CI
hamiltonian matrix nor the solution of the coupled-perturbed Hartree--Fock
equation for the derivatives of the MO coefficients are required.
We have implemented an interface to \texttt{MOLPRO} to
compute the integrals and nuclear forces from the $i$-FCIQMC density matrices.\cite{molpro}

As a first benchmark, we have applied the $i$-FCIQMC methodology to compute
the nuclear forces at several points along the dissociation curve of
molecular nitrogen, as the electronic wavefunction changes from single- to
strong multi-reference character.
\begin{figure}[h]
    \begin{center}
        \includegraphics[width=0.4\textwidth]{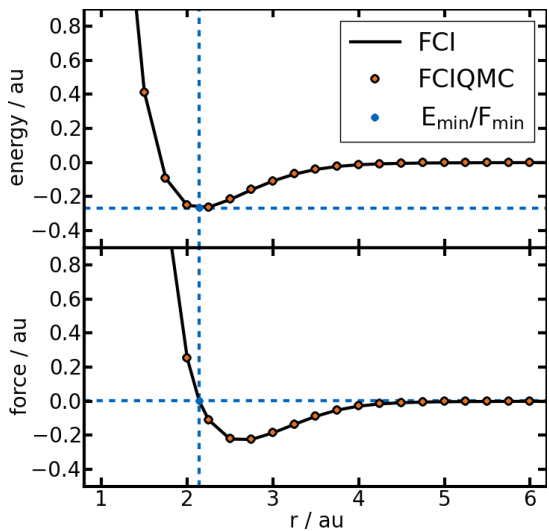}
    \end{center}
    \caption{Top: Potential energy profile for the N-N bond dissociation of
        N$_{\text{2}}$ relative to the energy of two isolated
        nitrogen atoms in the electronic ground state. 
        Bottom: corresponding forces at one nitrogen atom computed
        using analytic gradients from $i$-FCIQMC reduced density matrices, compared to FCI with numerical
        differentiation. Results are identical within the accuracy of the numerical differentiation.
        The respective minimum energy (E$_{\text{min}}$ = -0.2685\,a.u.)
        and force (F$_{\text{min}}$ = 0.0\,a.u.)
        at an internuclear distance of 2.144\,a.u. is indicated by the blue
        symbols.
        All results were obtained with a 6-31G basis set.}
    \label{fig:N2}
\end{figure}
Figure~\ref{fig:N2} (top) compares the potential energy computed with $i$-FCIQMC and the FCI
program in \texttt{MOLPRO} using a small \mbox{6-31G} basis set to allow for comparison to exact (FCI) results. 
The accuracy of the $i$-FCIQMC methodology for the computation of total
energies was already evaluated \cite{clelandboothalavi2010}, and we generally find
excellent agreement between the $i$-FCIQMC and FCI data set.

In Figure~\ref{fig:N2} (bottom), the nuclear forces for 
the same geometries are illustrated.
Comparison with FCI results obtained from numerical gradients provides a
direct measure of the quality of the reduced density matrices computed from
the replica algorithm based on $i$-FCIQMC, and, once again, the data shows excellent agreement between the analytic $i$-FCIQMC
forces and the FCI results for all geometries.

\begin{figure}[h]
    \begin{center}
        \includegraphics[width=0.4\textwidth]{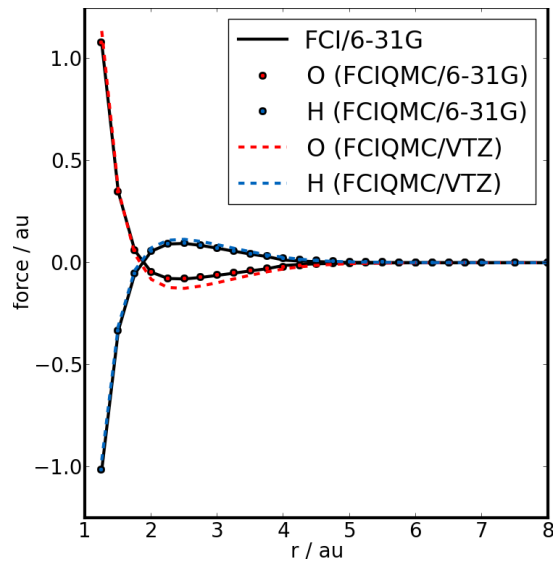}
    \end{center}
    \caption{Absolute forces acting on the oxygen and hydrogen atoms in a
        H$_{\text{2}}$O molecule computed using $i$-FCIQMC and FCI with a
        6-31G and cc-pVTZ basis set (the sign corresponds to the z-component of the force
        vector).
        The data were acquired for symmetric displacements of the
        hydrogen atoms from the equilibrium geometry. 
        The abscissa indicates the OH bond length of the respective
        molecular geometry.}
    \label{fig:H2O}
\end{figure}
As second example for the calculation of analytic gradients and
nuclear forces, we considered symmetric displacements of the atoms in
a water molecule along the OH bonds.
In a small 6-31G basis set, exact (FCI) diagonalisation of the hamiltonian
matrix is still feasible and Figure~\ref{fig:H2O} illustrates results from
FCI reference and $i$-FCIQMC calculations.
The nuclear forces as shown in Figure~\ref{fig:H2O} have been obtained
from the Cartesian force vectors as the absolute force acting on either a hydrogen
or the oxygen atom with the sign taken from the z-component of the force
vector, which has been aligned with the symmetry principal axis.
Although there is no computational advantage over direct diagonalisation
methods for basis sets as small as the 6-31G basis, the replica algorithm
implemented in $i$-FCIQMC can be applied to much larger molecules and basis
sets, providing essentially numerically exact nuclear forces.
In order to demonstrate the scope of the $i$-FCIQMC replica technology, we
have also computed the all-electron forces within a cc-pVTZ basis set, evidently an
infeasible task for current deterministic FCI algorithms, where the many-body basis now spans $\mathcal{O}[10^{13}]$ determinants. 
Figure~\ref{fig:H2O} (dashed lines) illustrates the notably larger
forces at intermediate stretching of the OH bonds if accurate 
cc-pVTZ basis set are combined with this level of theory in the calculations.
This would have implications for dynamics calculations, as well as providing the basis for 
highly accurate geometry optimisations for systems with
electronic ground states of strong multi-reference character.

\section{The dipole moment of CO}

The interaction of an electronic system of charge $q$ with an external electric field, $\bm{\xi}$, in an external potential, $V$, may be expressed as an expansion in terms of multipoles,
\begin{equation}
E = qV - \bm{\mu}\cdot\bm{\xi} - \frac{1}{2} \bm{\Theta}\cdot\frac{\partial\bm{\xi}}{\partial\mathbf{r}}-\ldots,
\end{equation}
with $\bm{\mu}$ the rank-1 dipole moment, $\bm{\Theta}$ the rank-2 quadrupole moment, and so on. It is the dipole moment itself with which we are presently concerned, and which may be calculated according to:
\begin{align}
\bm{\mu} &= \Braket{\Psi|\hat{\mu}|\Psi}\\
         &= \langle \Psi|\sum_i^N q_i\mathbf{r}_i|\Psi \rangle\\
         &= -\sum_i^N\Braket{\Psi|\mathbf{r}_i|\Psi},
\end{align}
where, in the last line, the substitution $q_i=-1$ (for electrons) has been made. Applying the Slater--Condon rules,\cite{slater1929,condon1930} this expression can be recast in terms of the one-body reduced density matrix and one-electron molecular-orbital integrals for an arbitrary Cartesian component, $w$, as,
\begin{equation}
\mu_w = -\sum_{pq} \gamma_{pq} \Braket{\phi_p|w|\phi_q} + \sum_I Z_I R_I^{\left(w\right)},
\end{equation}
to which the contribution from the (fixed) nuclei with charges $\left\{Z_I\right\}$ and positions $\left\{R_I\right\}$ has been added. Thus, given the molecular-orbital integrals, $\left\{\Braket{\phi_p|x|\phi_q}\right\}$, $\left\{\Braket{\phi_p|y|\phi_q}\right\}$, and $\left\{\Braket{\phi_p|z|\phi_q}\right\}$, which are readily available,\cite{hampelpetersonwerner1992,molpro} the one-body reduced density matrix obtained from $i$-FCIQMC provides direct access to the dipole moment, and, more generally, to multipole moments of arbitrary rank.

As an interesting application of this approach, we consider the well-known problem of the dipole moment of CO at its equilibrium bond length, $2.1316\,a_0$.\cite{huberherzberg1979} This system, with its subtle combination of $\sigma$ and $\pi$ effects, is difficult to predict intuitively \emph{a priori}, and Hartree--Fock theory notably suggests the polarity to be C$^+$O$^-$, while it is experimentally known to be C$^-$O$^+$.

We use the large aug-cc-pV$X$Z-DK basis sets for this study and adopt the second-order Douglas--Kroll--Hess hamiltonian.\cite{douglaskroll1974,hess1985,hess1986,balabanovpeterson2005,penghirao2009} Although relativistic effects are small for these comparatively light atoms, the calculation of the dipole moment tends to be strongly basis-set dependent, and the use of a large set becomes correspondingly desirable. To that same end, it is desirable to be able to extrapolate finite-basis dipole moments to the complete-basis-set limit, as such extrapolations have previously been useful in $i$-FCIQMC studies.\cite{thomasboothalavi2015} It has been shown that the asymptotic convergence of the correlation part of the dipole moment with the cardinality of the basis set, $X$, is suitably described by the form
\begin{equation}
\bm{\mu}_{\mathrm{corr}}^{\left(\mathrm{X}\right)}=\bm{\mu}_{\mathrm{corr}}^{\left(\mathrm{CBS}\right)} + \mathbf{a}X^{-3},
\end{equation}
in much the same way as the correlation energy itself.\cite{halkier1998,halkier1999} The complete-basis-set limit correlation contribution to the dipole moment, $\bm{\mu}_{\mathrm{corr}}^{\left(\mathrm{CBS}\right)}$, may thus be derived from two consecutive finite-basis results, of cardinality $X-1$ and $X$, according to
\begin{equation}
\bm{\mu}_{\mathrm{corr}}^{\left(\mathrm{CBS}\right)} = \frac{X^3\bm{\mu}_{\mathrm{corr}}^{\left(X\right)}- \left(X-1\right)^3\bm{\mu}_{\mathrm{corr}}^{\left(X-1\right)}}{X^3-\left(X-1\right)^3},
\end{equation}
to which the Hartree--Fock contribution in a suitably large basis (aug-cc-pV5Z-DK is used here, for which $\mu_{z\mathrm{,HF}}=-0.10355\,ea_0$) may then be added to obtain the total dipole moment.

\begin{table}
\begin{tabular}{cccccc}
               & \multicolumn{5}{c}{$\mu_z/ea_0$}\\
               &  \multicolumn{3}{c}{aug-cc-pV$X$Z-DK} & \multicolumn{2}{c}{CBS}\\
               & $X=$D & $X=$T & $X=$Q & (DT) & (TQ)\\
               \hline\hline
                 HF & -0.10135 & -0.10435 & -0.10369 &- &- \\
               MRCI & 0.07175 & 0.07203 & 0.07066 &0.07419 & 0.06929\\
               CCSD & 0.06829 & 0.05594 & 0.05087 &0.05278 & 0.04681\\
               $i$-FCIQMC & 0.05893(3)& 0.05200(4) & 0.0474(4)& 0.05112 & 0.0437\\
                \hline
\end{tabular} 
\caption{Calculated dipole moments, $\mu_z$, for CO at the HF, MRCI (using a $10$-electron, $8$-orbital active space),\cite{wernerknowles1988,knowleswerner1988,shamasundar2011} CCSD,\cite{hampelpetersonwerner1992} and \mbox{$i$-FCIQMC} levels of theory, with the complete-basis-set limit obtained from two-point, inverse-cube extrapolations.\cite{halkier1999} The standard error (in brackets) is derived as the standard deviation of the results from three independent \mbox{$i$-FCIQMC} calculations. The experimentally obtained bond length, $2.1316\,a_0$, is used,\cite{huberherzberg1979} and the $1\sigma^21\sigma^{*2}$ electrons are held frozen and neither relaxed nor optimised for the response of an electric field. The signs are arranged such that $\mu_z<0$ indicates a C$^+$O$^-$ polarity, and thus all the post-Hartree--Fock methods successfully reproduce qualitative agreement with the observed dipole's direction. The $i$-FCIQMC calculations were performed for $24$ hours on $400$ cores ($X=$D and T) or $600$ cores ($X=$Q)  using $\mathscr{O}\left(10^8\right)$ walkers, with the adjustable parameters $N_\mathrm{occ}=1$, $\chi=4$, $\kappa=0.01$, and $n_\mathrm{a}=3.0$, and the timstep allowed to vary dynamically to limit noisy walker growth. The sizes of the full orbital spaces for the double-, triple-, and quadruple-$\zeta$ calculations are $44$, $90$, and $158$, respectively.} 
\label{tab:CO-dipole}
\end{table}

\begin{figure}[h]
    \begin{center}
        \includegraphics[width=0.4\textwidth]{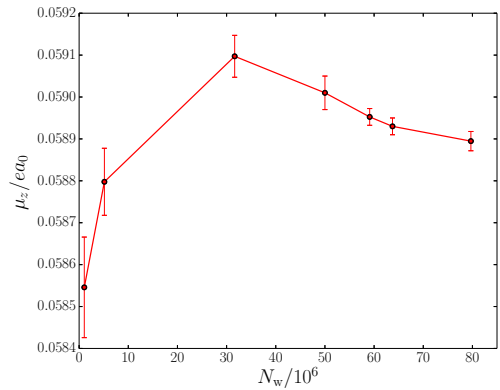}
    \end{center}
\caption{Calculated dipole moments for CO in an aug-cc-pVDZ-DK basis as a function of the number of walkers. Increasing the walker population is beneficial in reducing the stochastic error in the final result, but a qualitative description of the system is achieved at rather modest $N_\mathrm{w}$, as indicated by the fineness of the scale presented here.}    
\label{mu-conv}
\end{figure}

Table \ref{tab:CO-dipole} presents the results of this approach, with the extrapolations performed from the double- and triple-$\zeta$ and the triple- and quadruple-$\zeta$ basis sets, alongside 
the analogous results from coupled-cluster theory and multi-reference CI. The rapid convergence of $\mu_z$ with number of walkers in the $i$-FCIQMC dynamic is also demonstrated in Figure \ref{mu-conv}.

By comparison with the experimental dipole moment, variously given as $0.044\,ea_0$ and $0.048\,ea_0$,\cite{masonmcdaniel1988,muenter1975,scuseriamillerjensengeertsen1991,luis1995} it is apparent that $i$-FCIQMC performs rather better than MRCI, and is comparable to CCSD. 
However, it can be seen that CCSD actually overestimates the dipole moment compared to $i$-FCIQMC, which can be taken as close to exact in each of the finite basis sets, and 
this feature of CCSD allows for favourable cancellation of errors with the basis-set incompleteness, yielding the fortuitously accurate extrapolated result.
The remaining disparity between these results and experiment should not be ascribed to an inadequacy of the $i$-FCIQMC density matrices, but is rather largely attributable 
to basis-set incompleteness error. Indeed, the larger (TQ) extrapolations are rather more satisfactory than the corresponding (DT) results, highlighting the sensitivity of 
such approaches to the adequacy of the choice of basis. This effect has been previously observed in the context of ionisation potentials,\cite{thomasboothalavi2015} but is 
magnified in this instance by the stronger basis-set dependence of dipole moments than correlation energies. It is also worth noting that a small vibrational contribution to the 
dipole moment is expected,\cite{luis1997} but the results of this study support the view expounded in that work by Luis and coworkers, that an accurate treatment of electron 
correlation in a sufficiently large basis set is adequate for close agreement with experiment.

The quality of the $i$-FCIQMC density matrices may be illustrated by considering the CO problem in a small cc-pVDZ basis, for which deterministic FCI results can be obtained. In this case, whose results are summarised in Table \ref{tab:co-vdz}, $i$-FCIQMC reproduces the FCI dipole moment to within $0.06\%$, whilst the CCSD and MRCI results are in error by $10\%$ and $8\%$, respectively. Also of note is that, whilst the quoted $i$-FCIQMC result was obtained using $\mathscr{O}\left(10^8\right)$ walkers, it can be obtained just as well, and with apparently negligible initiator error, with only $\mathscr{O}\left(10^7\right)$.

\begin{table}
\begin{tabular}{ccc}
    & $\mu_z/ea_0$ & Abs. relative error (\%) \\
\hline\hline
HF   & -0.0915 & 201.10 \\
MRCI & 0.0973  & 7.61 \\
CCSD & 0.0996  & 9.94 \\
CCSDT & 0.0931 & 2.87\\
CCSDTQ & 0.0906 & 0.11\\
CCSDTQP & 0.0905 & - \\
$i$-FCIQMC & 0.09045(3) & 0.06 \\
FCI & 0.0905 & - \\
 \hline
\end{tabular}
\caption{Comparison of obtained dipole moments of CO in a small cc-pVDZ basis to the deterministic FCI result. As in Table \ref{tab:CO-dipole}, at all levels of theory, the two core orbitals were held frozen, and neither relaxed nor optimised for the response of an electric field. The $i$-FCIQMC result being in error by less than $0.1\%$, the density matrices derived therefrom are thus shown to be of near-FCI quality. The coupled-cluster results --- obtained by finite differentiation ($\pm 2\times 10^{-5}\,E_\mathrm{h}\,e^{-1}\,a_0^{-1}$) using the \texttt{MOLPRO}\cite{wernerknowles2012,molpro} and \texttt{MRCC}\cite{mrcc} codes --- are slow to converge to the FCI limit, with quadruple excitations needed for high accuracy.}
\label{tab:co-vdz}
    \end{table}

These results, therefore, bear out the supposed high quality of the sampled density matrices, and in demonstrating the compatibility of $i$-FCIQMC with the Hellmann--Feynman theorem, suggest that future studies of energy derivatives and their associated properties may well prove fruitful.

\section{Atomic dipole polarisabilities}

The previous section began by noting the dependence upon the permanent dipole moment of a system's interaction with an applied electric field as given by $-\bm{\mu}\cdot\bm{\xi}$. Of course, the application of such a field will, in reality, affect the distribution of charge, and hence the dipole moment itself. Expanding the dipole moment as a function of the field, therefore, we may write a given component, $\mu_i$, as
\begin{equation}
\mu_i = \mu_i^{\left(0\right)} + \sum_j \alpha_{ij}\xi_j + \frac{1}{2} \sum_{jk} \beta_{ijk}\xi_j\xi_k +\ldots,
\label{dipoleexpansion}
\end{equation}
where $\alpha_{ij}$ and $\beta_{ijk}$ represent elements of the polarisability and first hyperpolarisability tensors, respectively.\cite{greadybacskayhush1977} $\bm{\mu}^{\left(0\right)}$ is the zero-field permanent dipole, which is always zero for atomic species. Whilst the effect of the induced dipole moment is generally less significant for polar systems, it is the leading-order term in the expansion of the dipole moment for atoms which has no static dipole. It is thus crucial in accounting for the dipole-dipole dispersion interactions which often bind such species, and indeed will be the first-order response not only to static, but also to dynamic fields.\cite{stone} The calculation of $\bm{\alpha}$ thus provides an interesting study in and of itself, as well as a probing test of the calculation of reduced density matrices with $i$-FCIQMC. We here consider the noble-gas atoms, Ne, Ar, and Kr, as archetypal examples of the problem.

It is apparent from Eq.~\ref{dipoleexpansion} that the polarisability may be thought of as the derivative of the dipole with respect to the field,
\begin{equation}
\alpha_{ij} =\frac{\partial\mu_i}{\partial\xi_j} \bigg|_{\bm{\xi}=\mathbf{0}} ,
\end{equation}
evaluated at $\bm{\xi}=\mathbf{0}$. As for many response properties, this may be calculated by solution of the coupled perturbed Hartree--Fock equations,\cite{mcweeny} but for our purposes it is convenient to suppose that a particular component, $\alpha_{ij}$, might be effectively calculated by a finite-difference approach,
\begin{equation}
    \alpha_{ij} = \frac{\mu_i\left(\delta\xi_j\right) - \mu_i^{\left(0\right)}}{\delta\xi_j} = \frac{\mu_i\left(\delta\xi_j\right)}{\delta\xi_j},
\label{finitefield}
\end{equation}
in which $\delta\xi_j$ is a small field applied in the $j$ direction, and $\mu_i\left(\delta\xi_j\right)$ is the $i$th component of the dipole moment induced by so doing. The second equality holds for the spherically-symmetric atomic systems under consideration here since $\bm{\mu}^{\left(0\right)}=\mathbf{0}$, and the errors resulting formula are second-order since it is now equivalent to a central-difference approximation.

Straightforward and appealing though this implementation is, it is useful before proceeding to have some notion of its performance relative to analytic gradient methods. In particular, analytic gradients are readily and rapidly available from MP2 theory,\cite{hampelpetersonwerner1992} and this approach thus provides a useful framework in which to assess the suitability of the finite-field method.

\begin{table}
\begin{tabular}{cccc}
                     & \multicolumn{3}{c}{$\alpha_{zz}/e^2a_0^2E_\mathrm{h}^{-1}$}\\
                     & \multicolumn{3}{c}{aug-cc-pVTZ-DK}\\
              System & Analytic & Finite-field & Abs. relative error\\
\hline\hline
Ne & 2.438384 & 2.437962 & 1.73$\times 10^{-4}$\\
Ar & 10.841398 & 10.842952 & 1.43$\times 10^{-4}$ \\
Kr & 16.674939 & 16.680012 & 3.04$\times 10^{-4}$\\
\hline
\tiny{}\\
& \multicolumn{3}{c}{aug-cc-pVQZ-DK}\\
& Analytic & Finite-field & Abs. relative error\\
\hline\hline
Ne & 2.620174 & 2.619806 & 1.40$\times 10^{-4}$ \\
Ar & 11.128300 & 11.131348 & 2.74$\times 10^{-4}$ \\
Kr & 16.792296 & 16.799500 & 4.29$\times 10^{-4}$\\
\hline
\end{tabular}
    \caption[MP2-level comparison of the analytic and finite-field methods]{Analytic MP2 dipole polarisabilities, $\alpha_{zz}$, for the noble gases Ne, Ar, and Kr, in aug-cc-pVTZ-DK and aug-cc-pVQZ-DK basis sets compared with the corresponding finite-field results, calculated with an electric field strength of $0.005$ $E_{\mathrm{h}}/ea_0$.}
    \label{tab:analytic}
    \end{table}

\begin{figure}[h]
    \begin{center}
        \includegraphics[width=0.4\textwidth]{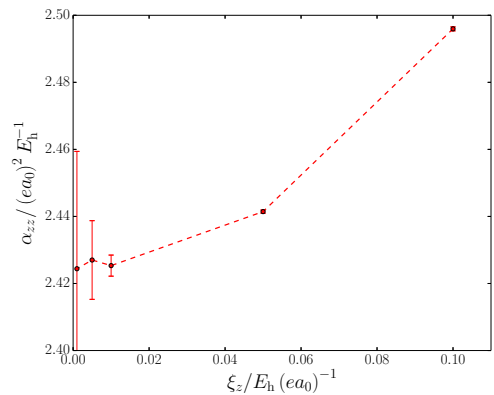}
    \end{center}
    \caption{Calculated dipole polarisabilities for the Ne atom in an aug-cc-pVTZ-DK basis with different applied field strengths, $\xi_z$. As in the previous section, the $i$-FCIQMC calculations were performed using $\mathscr{O}\left(10^8\right)$
walkers, a dynamic timestep, and the adjustable parameters $N_\mathrm{occ}=1$, $\chi=4$, $\kappa=0.
01$, and $n_\mathrm{a}=3.0$. Sufficiently small fields establish the required pseudo-linear relationship between the polarisability and the applied field, but too small a field gives rise to large stochastic errors. This initial study prioritises the elimination of non-linear effects, and a field strength of $\xi_z=0.005\,E_\mathrm{h}/ea_0$ is thus chosen as suitable for the remainder of this work, where the random errors can be systematically controlled. It is encouraging to note, however, that this choice may be somewhat conservative, and that a slightly larger field may be permissible in future work.}
    \label{fieldstrength}
\end{figure}

The results of this comparison, with the finite-field polarisabilities performed in a field of strength 0.005 $E_{\mathrm{h}}/ea_0$, are summarised in Table \ref{tab:analytic}. The mean absolute percentage error inherent in the approach is found to be of the order of $0.02\%$, demonstrating both its suitability for the purpose, and also that the field chosen is sufficiently small to establish the pseudo-linear dependence of the induced dipole upon the field. That this dependence is established without having to use a very small field is encouraging, since in the stochastic formulation provided by $i$-FCIQMC, the stochastic error in the induced dipole must be divided by the field strength to obtain the equivalent error bounds in the polarisability. This behaviour is illustrated for Ne in the aug-cc-VTZ-DK basis in Figure \ref{fieldstrength}, which highlights the balance which must be achieved between minimising second-order effects and maintaining a suitable level of stochastic error.

Secure in the knowledge of the suitability of the finite-field approach, we may proceed with an assessment of the performance of $i$-FCIQMC compared with other methods. Specifically, as in the previous section, we calculate the polarisabilities using CCSD and MRCI for comparison,\cite{hampelpetersonwerner1992,wernerknowles1988,knowleswerner1988,shamasundar2011} though in this case the extrapolations are performed from results at the triple- and quadruple-$\zeta$ basis sizes, reflecting the increased sensitivity to basis set incompleteness error which this quantity entails.

\begin{table}
\begin{tabular}{ccccc}
                     & \multicolumn{4}{c}{$\alpha_{zz}/e^2a_0^2E_\mathrm{h}^{-1}$}\\
System & aug-cc-pVTZ-DK & aug-cc-pVQZ-DK & CBS & Experiment \\
\hline\hline
Ne & 2.42(1)& 2.596(2) & 2.65 & 2.57 \\
Ar & 10.855(5) & 11.092(3) & 11.08 & 11.23\\
Kr & 16.81(4) & 16.86(6) & 16.82 & 16.73 \\
 \hline
\end{tabular}
    \caption[$i$-FCIQMC polarisabilities of the noble gases]{$i$-FCIQMC polarisabilities of the noble gases Ne, Ar, and Kr, obtained in aug-cc-pVTZ-DK and aug-cc-pVQZ-DK basis sets, along with the extrapolated complete-basis-set limit results. The number in brackets indicates the error in the preceding digit, obtained as the standard deviation of the results of three independent calculations. The experimental results are shown for comparison.\cite{delgarno1960,olney1997} The $i$-FCIQMC calculations were performed using $\mathscr{O}\left(10^8\right)$
walkers, the adjustable parameters $N_\mathrm{occ}=1$, $\chi=4$, $\kappa=0.
01$, and $n_\mathrm{a}=3.0$, and run for $48$ hours on $320$ cores.}
        \label{tab:polarisabilities}
    \end{table}

As might have been expected, the error incurred by extrapolating is somewhat reduced upon including the larger quadruple-$\zeta$ treatment, and the $i$-FCIQMC results given in Table \ref{tab:polarisabilities} bear correspondingly close agreement with experiment. The remaining errors --- in the region of $0.5$ to $3\%$ --- are nonetheless still likely to be artefacts of the basis sets, as the application of a field accentuates the importance of describing the intricacies of the more diffuse regions of electron density. Thus, although ``augmented'' basis sets are employed, there is likely still something to be gained from a more complete description of this behaviour.
This suggestion is, once again, further strengthened by the fact that $i$-FCIQMC is able to recover FCI-quality results for basis sets in which direct comparison is possible, reproducing the polarisability of Ne in a small cc-pVDZ basis to within $0.005\%$, for instance.

The same results, computed using Hartree--Fock theory, CCSD,\cite{hampelpetersonwerner1992} and MRCI,\cite{wernerknowles1988,knowleswerner1988,shamasundar2011} are listed in Table \ref{tab:other}. The mean (absolute) error for the MRCI calculations is $4.7\%$, whilst that for $i$-FCIQMC, and coupled-cluster theory, is around $1.6\%$. The comparability is unsurprising, given the ascription of much of the error to finite-basis effects. However, it is now necessary to investigate the impact of stronger correlation on this quantity in more challenging systems, where we expect more significant advantages to come from $i$-FCIQMC. 

\begin{table}
\begin{tabular}{ccccccccc}
& \multicolumn{8}{c}{$\alpha_{zz}/e^2a_0^2E_\mathrm{h}^{-1}$}\\
& \multicolumn{3}{c}{aug-cc-pVTZ-DK} & \multicolumn{3}{c}{aug-cc-pVQZ-DK} & \multicolumn{2}{c}{CBS}  \\
System & HF & CCSD & MRCI & HF & CCSD & MRCI & CCSD & MRCI \\
\hline\hline
Ne & 2.20 & 2.42 & 2.41 & 2.33 & 2.59 & 2.58 & 2.64 & 2.64 \\
Ar & 10.45 & 10.81 & 10.36 & 10.72 & 11.03 & 11.53 & 11.00 & 12.20 \\
Kr & 16.21 & 16.78 & 17.15 & 16.36 & 16.81 & 17.22 & 16.74 & 17.18\\
 \hline
\end{tabular}
\caption[Noble-gas polarisabilities at various levels of theory]{Polarisabilities of the noble gases Ne, Ar, and Kr, computed using Hartree--Fock, coupled-cluster, and multi-reference CI (with an $8$-electron, $8$-orbital active space) theories. As in Table \ref{tab:polarisabilities}, the extrapolations to the complete-basis-set limits are also shown.\cite{delgarno1960,olney1997}}
\label{tab:other}
    \end{table}

\section{Conclusions}

The results presented in this work serve to confirm the high quality of the stochastically-obtained reduced density matrices available \emph{via} replica sampling in $i$-FCIQMC, capable as they are of reproducing FCI-quality results for nuclear forces, dipole moments, and polarisabilities, and in some cases close agreement with experimental values. In so doing, they cement the place of the replica technique as an important extension to the theory, and widen its scope considerably.

In addition to the most obvious extension of an ability to compute a larger range of properties for a wider variety of systems, there remain a number of theoretical and technical challenges to be addressed in future studies. Perhaps the most pressing task is to extend this work to encompass results from open-shell systems, in which correlation effects are likely to be more important. Moreover, if comparisons to experimental results are to be further sought and achieved for dipole moment properties, there is some motivation to explore larger basis sets with multiple levels of augmentation,\cite{kendall1992,woon1993} which may be of particular use in better describing the more diffuse electron densities of finite-field calculations, and more generally in describing larger and heavier atoms of interest.

\section{Acknowledgements}

The authors would like to thank Gerald Knizia for helpful discussions. The authors gratefully acknowledge Trinity College, Cambridge, and the Royal Society \emph{via} a University Research Fellowship for funding.  This work was also supported through a research fellowship of the Deutsche Forschungsgemeinschaft (D.\,O.) and by EPSRC under Grant No. EP/J003867/1. The calculations made use of the facilities of the Max Planck Society's Rechenzentrum Garching.


%

\end{document}